\documentclass[aps,amssymb,pre,twocolumn,showpacs,superscriptaddress]{revtex4}

\usepackage[english]{babel}
\usepackage[latin1]{inputenc}
\usepackage[dvips]{graphicx}
\usepackage{amsmath}
\usepackage{epsfig}
\usepackage{graphicx,color}
\usepackage{array}

\newcommand{\beq}{\begin{equation}}
\newcommand{\eeq}{\end{equation}}
\newcommand{\beqa}{\begin{eqnarray}}
\newcommand{\eeqa}{\end{eqnarray}}

\newcommand{\fn}{\varphi_{\text{nem}}}
\newcommand{\Csin}{\varphi^{(I)}_{s}}
\newcommand{\Cbin}{\varphi^{(I)}_{b}}
\newcommand{\Csni}{\varphi^{(N)}_{s}}
\newcommand{\Cbni}{\varphi^{(N)}_{b}}
\newcommand{\ti}{\tau_{\text{ind}}}

\newcommand{\eq}[1]{Eq.~(\ref{#1})}
\newcommand{\fig}[1]{Fig.~\ref{#1}}
\newcommand{\bhu}{{\bf \hat{u}}}

\newcommand{\olcites}[1]{Refs.~\onlinecite{#1}}

\begin{document}
\bibliographystyle{apsrev}

\title{Supersaturated dispersions of rod-like viruses with added attraction}
\author{ P. Holmqvist}
\affiliation{Institut f\"ur Festk\"orperforschung, Forschungszentrum
J\"ulich, D-52425 J\"ulich, Germany}

\author{ M. Ratajczyk}
\affiliation{Institut f\"ur Festk\"orperforschung, Forschungszentrum
J\"ulich, D-52425 J\"ulich, Germany}
\affiliation{Institute of Physics, A. Mickiewicz University, Umultowska 85, 61-614, Poznan, Poland}

\author{ G. Meier}
\affiliation{Institut f\"ur Festk\"orperforschung, Forschungszentrum
J\"ulich, D-52425 J\"ulich, Germany}

\author{H. H. Wensink}
\affiliation{Department of Chemical Engineering, Imperial College London, South Kensington Campus, London SW7 2AZ, United Kingdom}

\author{M. P. Lettinga}
\affiliation{Institut f\"ur Festk\"orperforschung, Forschungszentrum
J\"ulich, D-52425 J\"ulich, Germany}

\date{\today}

\begin{abstract}
The kinetics of isotropic-nematic (I-N) and nematic-isotropic (N-I) phase transitions in dispersions of rod-like {\it fd}-viruses are studied. Concentration quenches were applied using pressure jumps in combination with polarization microscopy, birefringence and turbidity measurements. The full biphasic region could be accessed, resulting in the construction of a first experimental analogue of the bifurcation diagram. The N-I spinodal points for dispersions of rods with varying concentrations of depletion agents (dextran) were obtained from orientation quenches, using cessation of shear flow in combination with small angle light scattering. We found that the location of the N-I spinodal point is independent of the attraction, which was confirmed by theoretical calculations. Surprisingly, the experiments showed that also the absolute induction time, the critical nucleus and the
growth rate are insensitive of the attraction, when the concentration is scaled to the distance to the phase boundaries.
\end{abstract}

\pacs{82.70.Dd, 64.70.Md, 82.60.Lf, 83.80.Xz}

\maketitle
\section{Introduction}

A long standing issue in the physics of fluids is the behavior of the homogeneous fluid close to the point where it becomes unstable and phase separates, i.e. the \textit{spinodal} point $\varphi_{s}$~\cite{Debenedetti96}. Before the spinodal point the fluid will be meta-stable or supersaturated, which means that the fluid will only undergo a phase transition when fluctuations in the concentration are sufficiently high to overcome a certain nucleation barrier. Thus the meta-stable region is characterized by the induction time $\ti$ for phase separation to set in. $\ti$ goes to infinity entering the meta-stable region from the stable region at the \emph{binodal} point $\varphi_{b}$ , i.e. $1/\ti \rightarrow 0$ at $\varphi_{b}$, while $\ti  \rightarrow 0$ approaching $\varphi_{s}$. For molecular fluids it is very difficult determine the spinodal point, because the tiniest impurity will lower the nucleation barrier and the phase separation is very fast. The binodal point is easier to access, since it is given by the final phase separated state. Colloidal systems have proven to be very suitable for this type of fundamental studies. The main reason is that the interactions between the colloids can be tailored, while the size of the colloids slows down the kinetics as compared to fluids permitting direct visualization~\cite{Anderson02}. One way of tailoring the interaction between colloids is to add non-adsorbing polymers to the system. Polymers induce attractive interaction between the colloids, due to the depletion of the polymer between the colloids~\cite{Asakura58}. The range and the strength of the attractive potential is controlled by the polymer size and concentration, respectively. Due to attractions colloid-polymer mixtures typically show a gas-liquid-like phase transition~\cite{Lekkerkerker92}. Despite of the advantages of colloids, it is also for this class of systems difficult to access the meta-stable and unstable region in a controlled way. Firstly, the spinodal and binodal line are located very close to each other, while the energy barrier for phase separation to take place is low. Secondly, systems might be arrested in the metastable state\cite{Verduin95b}. The challenge is to bring the system in a meta-stable state while maintaining the system homogeneous.

\begin{figure}
\includegraphics[width=.4\textwidth]{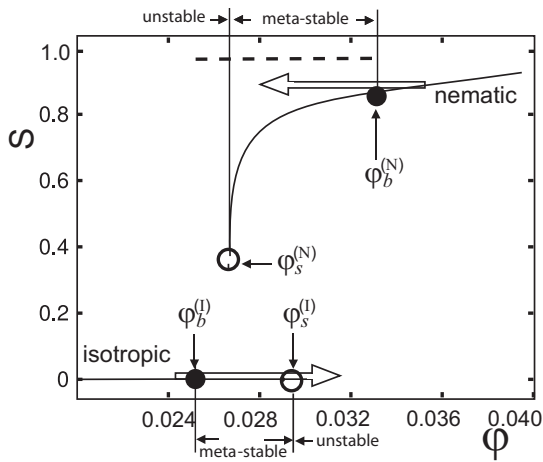}
\caption{Bifurcation diagram of the nematic order parameter $S$ for hard rods ($\varphi_{p}^{R}=0$) with $L/D=133$ corresponding to the free energy \eq{free}. Open circles indicate spinodal points, while the filled circles indicate binodal points. The arrows indicate concentration quenches that are made to render the system in a supersaturated state. The dashed line indicates possible locations of the system when inducing a full nematic phase applying an external field.} \label{fig:PhaseScheme}
\end{figure}

Where concentration is the only order parameter of interest for gas-liquid phase separating colloidal spheres, dispersions of colloidal rods have two order parameters that characterize the phase behavior which are strongly coupled: particle concentration and orientational order. These systems exhibit an isotropic-nematic (I-N) phase transition, where the system gains positional entropy at the cost of the orientational entropy. The location of the transition can be derived from Onsager theory for slender hard rods~\cite{Onsager49}. The phase behavior around the I-N transition is characterized by two branches~\cite{Kayser78}: an isotropic branch with zero orientational order and a nematic branch with a finite orientational order, as depicted in Fig. \ref{fig:PhaseScheme}.
The I-N binodal point $\varphi_{b}^{(I)}$ will be first encountered when following the isotropic branch by increasing the concentration. For $\varphi>\varphi_{b}^{(I)}$ the system will be meta-stable, or supercooled, with respect to fluctuations in the orientation towards an aligned state. After some typical induction time $\ti$ fluctuations will be sufficient to overcome the nucleation barrier and isolated nematic droplets (tactoids) will grow in a isotropic background. For even higher concentrations, i.e. for  $\varphi>\varphi_{s}^{(I)}$  the system becomes unstable to fluctuations, such that each fluctuation in the orientation of the rods will result in a continuous growth of the nematic phase out of the isotropic phase. Likewise, the system will have a N-I binodal $\varphi_{b}^{(N)}$ and spinodal $\varphi_{s}^{(N)}$ point when following the nematic branch by decreasing the concentration. In this case the system is superheated because fluctuations towards a lower ordering  drive the phase separation. Beyond the binodal point, i.e. for $\varphi<\varphi_{b}^{(N)}$ isotropic nuclei (atactoids) will form after some induction time $\ti$, while beyond the spinodal point, i.e. for $\varphi<\varphi_{s}^{(N)}$,  each fluctuation in the orientation of the rods initiates spinodal phase separation and an isotropic phase of disordered rods continuously grows out of the nematic phase.

The difference in concentration between the I-N and N-I binodal points of a hard rod liquid is only about $10\;\%$~\cite{Vroege92}. Hence the spinodal and binodal points are located very close to each other. The density difference between the binodal points increases when the rods are made attractive: the I-N binodal shifts to lower concentrations while the N-I binodal shifts to higher concentrations. This has been shown experimentally for various rod-polymer mixtures~\cite{Dogic04a,vanbruggen00,Buitenhuis95,Edgar02}. It is, however, not obvious that the location of the spinodal points are equally affected by adding polymer.
As a consequence it is not known to what extend the metastable and unstable regions are affected by the addition of polymer. Even for hard rods the location of the spinodal points, i.e. the open symbols in Fig. \ref{fig:PhaseScheme}, has never been experimentally confirmed. The goal of this paper is to locate the spinodal points  with respect to the binodal points over a range of attractions by probing the kinetics of the phase separation process. This goal requires a well defined time $t=0$ at which the system is quenched from an initially stable state into a meta-stable or unstable state, in order to determine the induction time $\ti$. Taking advantage of the two order parameters that characterize rod dispersions also two types of quenches can be made: a quench in the orientation and in concentration.

An orientation quench can be  performed by first applying an external field to a phase separated system somewhere in the biphasic region, thus preparing a full nematic phase, as indicated by the dashed line in Fig. \ref{fig:PhaseScheme}. The system is quenched when at $t=0$ the external field is switched off, which renders the system either unstable or meta-stable depending on the concentration. Tang and Fraden used the diamagnetic anisotropy of {\it fd} virus to induce I-N phase transitions with a high magnetic field~\cite{Tang93} and demonstrated nicely the existence of an unstable region. Here they quenched, however, always to some finite field strength. Similarly, we used in an earlier paper shear flow to prepare a stable nematic phase~\cite{Lettinga05c}. Cessation of flow at $t=0$ renders the nematic phase meta-stable or unstable depending on the concentration, see Fig. \ref{fig:PhaseScheme}. Small Angle Light Scattering (SALS) was then used to probe the formed biphasic structures and determine the spinodal as the concentration where structure formation sets in immediately after the quench. In this paper we again rely on this technique on mixtures of \emph{fd} and dextran, but as compared to the earlier experiments the sensitivity is improved such that measurements could be performed also when density differences between the phases were small as is the case at low polymer concentrations.

The disadvantage of the orientation quench is that only the nematic-isotropic transition is probed. In order to access also the I-N transition one needs to make a concentration quench. Such a quench can be made by rigorous stirring a phase separated system and probe it immediately after stirring, as was done for dispersions of boehmite rods~\cite{vanbruggen99a}.

Apart from the practical problems this imposes on the experiment, the results could also be biased by residual alignment in the sample after mixing~\cite{ripoll08}. Quenches were also initiated by polymerizing short actin filaments~\cite{Oakes07}. Both methods could evidence the distinction between nucleation-and-growth and spinodal decomposition by the morphology of the phase separated structures, without pinpointing the actual location of the spinodal points. Nucleation-growth mechanisms and spinodal structures have also been observed in computer simulations~\cite{Schilling04,Cuetos07,Cuetos08,Berardi07}. In this paper we perform pressure quenches from 1 bar up to 1000 bar and vise versa. Given the compressibility of water, this corresponds with instantaneous concentration quenches of up to $5\;\%$~\cite{Fine73}, as indicated by the solid arrows in Fig. \ref{fig:PhaseScheme}. We probe changes using polarization microscopy, birefringence and turbidity measurements. We could determine both the isotropic to nematic (I-N) and nematic to isotropic (N-I) spinodal, since full nematic phase could be induced, starting with a full isotropic phase. Thus we construct a first experimental analogue of the bifurcation diagram plotted in Fig. \ref{fig:PhaseScheme}. Since with pressure quenches only a small concentration range can be accessed, we relied on cessation of shear flow to study the attractive rods with added dextran.

To supplement the experiments we have used Scaled Particle Theory (SPT) approach to predict the phase behavior of colloidal rods for a range of polymer concentrations, including the I-N and I-N spinodal lines. The experimental data are qualitatively compared with this theory. In Sec. \ref{sec:theory} we introduce the SPT and present its results in the form of two phase diagrams. In Sec. \ref{sec:Experimental} we introduce our experimental techniques and sample preparation. The effect of the pressure and orientation quenches are given in Sec. \ref{ssec:ExpPres} and \ref{ssec:ExpShear}, respectively, resulting in the determination of the spinodal and binodal points in Sec. \ref{ssec:ResSpBi} and growth rates in Sec. \ref{ssec:Grow}.

\section{Theory}\label{sec:theory}

The phase diagram of a rod-polymer mixture can be predicted from
free-volume theory, as elaborated in detail in
\olcites{lekkerkerker94,Tuinier07}. The free energy per
particle of a system of $N$ hard spherocylindrical rods with length $L$
and diameter $D$ in a volume $V$ in osmotic equilibrium with a reservoir
of ideal, non-adsorbing polymer with a volume fraction $\varphi_{p}^{R}$
takes the following form:
\beq
\frac{\beta F}{N} \sim \log y + \sigma[f] + P[f]y +
\frac{1}{2}Q[f]y^{2} - \frac{(3\gamma - 1) \varphi_{p}^{R}}{2q^3}\frac{\alpha
([f],\varphi ) }{\varphi} \label{free}
\eeq
in terms of the thermal energy $\beta^{-1}=k_{B}T $, rod aspect ratio $\gamma = L/D \gg1 $ and polymer-colloid
size ratio $q=2R_{g}/D$  (with $R_g$ the polymer radius of gyration).
The density variable $y=\varphi/(1-\varphi)$ is related to the rod packing
fraction $\varphi=(\pi/4)LD^{2}N/V$. The reference part $\varphi_{p}^{R}=0$
corresponds to a system of hard rods and stems from Scaled Particle Theory
(SPT). The orientational entropic contribution in \eq{free} is
defined as \beq
\sigma [f] = \int d \bhu f(\bhu) \ln [ 4 \pi f(\bhu)] \label{sigma} \eeq

\noindent where the unspecified distribution $f(\bhu)$, describes the probability of  rods
with orientational unit vector $\bhu$,
normalized over all possible orientations via $\int d \bhu f(\bhu) =1$. The coefficients $P$ and $Q$ pertain to
the shape (i.e. aspect ratio) of the rods:
\beqa
P[f] &=& 3 + \frac{3(\gamma
- 1)^{2}}{3 \gamma - 1} \tau[f] \nonumber \\ Q[f] &=&
\frac{12\gamma(2\gamma-1)}{(3\gamma -1 )^2} + \frac{12 \gamma (\gamma -
1)^2}{(3\gamma - 1)^2} \tau[f]
\eeqa
The quantity $\tau[f]$ represents the following double orientational
average: \beq \tau [f] = \frac{4}{\pi} \iint d \bhu d \bhu{^{\prime}}
f(\bhu) f(\bhu ^{\prime}) | \bhu \times \bhu^{\prime}| \label{tau} \eeq

The last term in \eq{free} accounts for the depletion contribution.
It depends  on the {\em free volume fraction} $\alpha$,
expressing the average fraction of the system volume available to the
polymer at a given rod packing fraction $\varphi$. An explicit expression
follows  from SPT: \beq \alpha ([f], \varphi) = (1-\varphi) \exp \left ( -
A y - B[f] y^{2} - C[f] y^{3} \right ) \eeq with coefficients $A,B,C$
given explicitly in ~\cite{Tuinier07}. Since the reservoir polymer
concentration $\varphi_{p}^{R}$ is proportional to the depth of the
minimum of the attractive {\em
depletion potential}, it serves as a measure for the
strength of attraction between the rods. As the polymers are treated
as an ideal gas, the
polymer volume fraction $\varphi_{\text{poly}}$ in the {\em system}
 simply follows from  multiplying the reservoir value  $
\varphi_{p}^{R}$ with the fraction of available free volume  $
\alpha([f], \varphi)$.

The SPT coefficients $P$, $Q$ and free volume fraction depend implicitly
on the unspecified orientational distribution $f(\bhu)$ . In the isotropic state, all orientations are equally probable so that
$f=1/4\pi$, $\sigma \equiv 0$ and $\tau \equiv 1$. In the nematic state,
it will be a  non-uniform distribution peaked along some nematic
director ${\bf \hat{n}}$. An accurate variational form for $f$ has been
proposed by
Onsager ~\cite{Onsager49} which takes the following form: \beq f(\theta) =
\frac{\kappa \cosh ( \kappa \cos \theta ) }{4 \pi \sinh \kappa }
\label{ons} \eeq with $0 \le \theta \le \pi $ the polar angle between
$\bhu$ and the nematic director ${\bf \hat{n}}$ ($\cos \theta = \bhu \cdot
{\bf \hat{n}}$) and $\kappa \ge 0 $ a variational order parameter (note
that $\kappa=0$ leads back to the isotropic constant $f=1/4\pi$). With the
use of an explicit trial function, the orientational averages
associated with \eq{tau} and \eq{sigma} become analytically tractable
~\cite{Onsager49,francomelgar08}. The results for the orientational averages
are: \beq \sigma(\kappa) = \ln (\kappa \coth \kappa) - 1 + \frac{\arctan
(\sinh \kappa)}{\sinh \kappa} \ge 0 \eeq and \beq \tau(\kappa) = \frac{2
I_{2}(2 \kappa)}{2 \sinh ^{2} \kappa} \le 1 \eeq with $I_{2}(x)$ a
modified Bessel function.

The equilibrium value for $\kappa$ is found by a minimization of the total free energy which leads to the stationarity
condition:
\beq \frac{\partial F}{\partial \kappa} \equiv 0 \label{stat}
\eeq

\noindent for any given rod packing fraction and attraction strength
$\varphi_{p}^{R}$. The nematic order parameter associated with the
equilibrium value for $\kappa$ is found from: \beqa S &=& \int d \bhu
{\cal P}_{2} (\bhu \cdot {\bf \hat{n}}) f(\bhu) \nonumber \\
  &=& 1- \frac{3 \coth \kappa}{\kappa} + \frac{3}{\kappa^{2}} \eeqa where
$S \equiv0$ in the isotropic and $0<S<1$ in the nematic phase.
The solution of \eq{stat} for hard rods ($\varphi_{p}^{R}=0$) is given in Fig. \ref{fig:PhaseScheme}, showing two  branches where the stationary solutions correspond to a local minimum of the free energy~\cite{Kayser78}. The spinodal points marks the transition between a stable and an unstable solution of \eq{stat}.

However, this (second order) transition is preempted by a first order phase transition. Thus a discontinuity both in concentration and in the ordering of the system will occur. The co-existence of two phases requires that the osmotic pressure $\Pi$ and chemical potential $\mu$ of the isotropic phase, with volume fraction $\varphi_{b}^{(I)}$, and the nematic phase, with volume fraction $\varphi_{b}^{(N)}$, are equal:
\beq \Pi(\varphi_{b}^{(I)})\;=\;\Pi(\varphi_{b}^{(N)})\eeq
\beq \mu(\varphi_{b}^{(I)})\;=\;\mu(\varphi_{b}^{(N)})\eeq
The binodal points can thus be found using the thermodynamic relations $\mu=\frac{\partial F}{\partial N}_{V,T}$ and $\Pi=-\frac{\partial F}{\partial V}_{N,T}$  in combination with Eq. \ref{free}.

\begin{figure}
\includegraphics[width= \columnwidth]{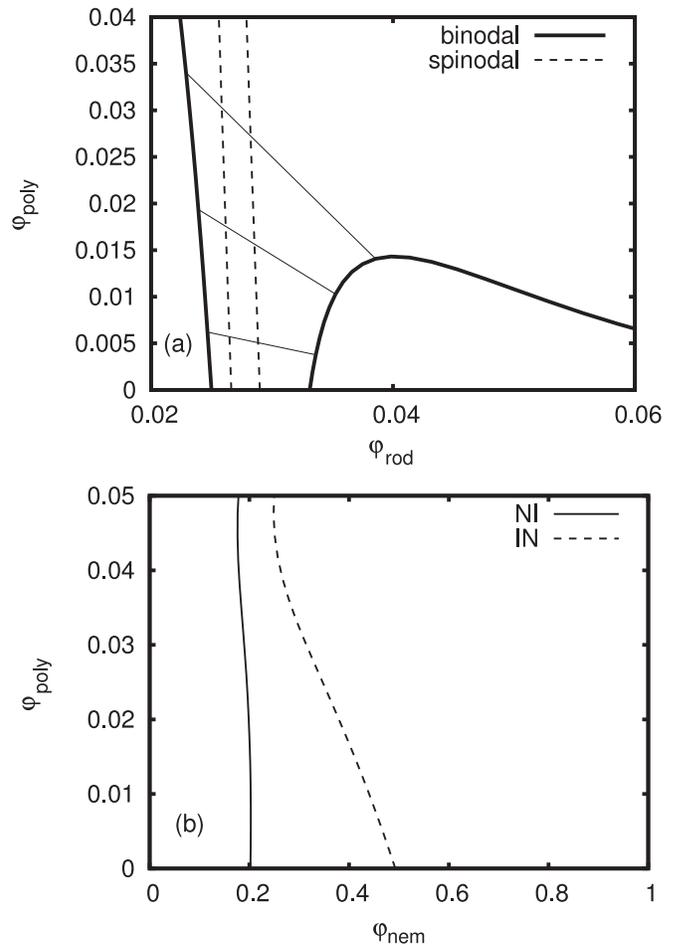}
\caption{ (a) Phase diagram for a rod-polymer mixture with $L/D=133$ and colloid-polymer size ratio $q=2R_{g}/D=5.4$.  Plotted in terms of the rod packing fraction $\varphi_{\text{rod}}$ and polymer volume fraction $\varphi_{\text{poly}}$ in the {\em system}. Coexisting isotropic and nematic phases are connected by tie lines, with the nematic phase  having a higher $\varphi_{\text{rod}}$. (b) Location of the isotropic-nematic spinodals in terms of the fraction of nematic phase $\varphi_{\text{nem}}$ [\eq{phi}] plotted versus the system polymer concentration $\varphi_{\text{poly}}$  on the vertical axis. \label{fig:theo1}}
\end{figure}

The phase diagram for a rod-polymer mixture is given in Fig. \ref{fig:theo1}a and shows the
characteristic widening of the biphasic gap as the amount of polymer is
increased. The location of the spinodal points, however, appears much less
affected by the depletion attraction. This is reflected more
clearly if we plot the spinodal curves in terms of the fraction $ 0 \le
\varphi_{\text{nem}} \le 1$ of nematic phase formed, rather than the overall rod
packing fraction $\varphi$. Applying the lever rule, we may relate
$\varphi_{\text{nem}}$ to $\varphi$ via:

\beq \varphi_{\text{nem}} = \frac{ \varphi -
\varphi_{b}^{(I)}}{\varphi_{b}^{(N)} - \varphi_{b}^{(I)}} \label{phi} \eeq

\noindent  with $\varphi_{b}^{(I/N)}$ the binodal rod packing fractions corresponding to the
coexisting isotropic and nematic phases. \fig{fig:theo1}b shows that the NI
spinodal instability occurs if the overall rod concentration corresponds
to the nematic phase occupying about 20 \% of the system volume. This
result is virtually independent of the strength of the depletion
attraction as long as the amount of added polymer is not too large.

\section{Experimental}\label{sec:Experimental}

\subsection{Sample}\label{ssec:Sample}

\begin{figure}
\includegraphics[width=.4\textwidth]{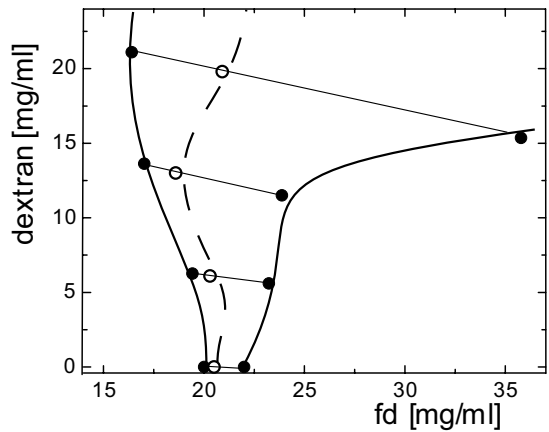}
\caption{Phase diagram of the I-N transition of $[dextran]$ vs. $[fd]$ at an ionic strength of $110\; mM$. The solid symbols are the binodal points as determined spectroscopically after phase separation. The open symbols are the spinodal points as determined after shear rate quenches, see below. The thin lines connecting the binodal points are the tie lines.} \label{fig:PhaseDex}
\end{figure}

{\it Fd}-virus suspensions were used in a 20 mM Tris buffer with 100 mM NaCl at a pH of 8.2. The virus is a long and thin rod like particle (length 880 nm long, width 6.6 nm, persistence length $2.2 \mu m$). Attractions between the rods are varied through depletion by addition of dextran (480 kd, Pharmacosmos). A small amount of FITC-labeled dextran was added to be able to determine the dextran concentration spectroscopically. See Ref. \cite{Tromp01} for the labeling procedure of dextran. The samples were prepared as follows: First, a homogeneous fd-virus suspension of 21.1 mg/ml \emph{fd }virus with dextran is allowed to macroscopically phase separate into an isotropic and nematic phase. The two phases were then separated into two different vials and the dextran and fd-virus concentrations were determined spectroscopically. Three different dextran concentration were used in this study  given an initial virus concentration of 21.1 mg/ml: 6 mg/ml (low), 13 mg/ml (middle) and 20 mg/ml (high). The resulting phase diagram is shown in Fig. \ref{fig:PhaseDex}. This phase behavior differs somewhat from previous published results for the same system~\cite{Dogic04a,Lettinga05c}. The deviation might be due to different polydispersity of the dextran which can drastically change the interaction~\cite{Kleshchanok06} and thus the phase behaviour\cite{Tuinier03}.  By combining different volumes of the isotropic, $\varphi_{b}^{(I)}$, and the nematic, $\varphi_{b}^{(N)}$, phase from the initially phase separated sample we can prepare any concentration along one tie-line, with a concentration $\fn$ relative to the phase boundaries as expressed in Eq. \ref{phi}. The concentration of dextran and fd-virus for each sample was checked after every new mixing. For pressure experiments a sample very close to the isotropic-nematic spinodal has been prepared at the same ionic strength, but without dextran.

\subsection{Microscopy, Birefringence, and turbidity at high pressure}\label{ssec:ExpPres}

For all pressure experiments we used a small container sealed by a vitron ring, which contains the sample while it allows for pressurizing via holes in the brass support ring. For microscopy and birefringence measurements the container was placed in the polarization microscopy cell, where the polarization was maintained. The cell was mounted in a specially designed cell holders for microscopy or birefringence.  For turbidity measurements a SANS cell was used which has longer optical pathway, to increase the sensitivity of the experiments, but which has windows that scramble the polarization. The detailed description of the cell and the preparation procedure can be found in Ref. ~\cite{Kohlbrecher07}.

Polarization microscopy was performed on a Zeiss Axioplan microscope
with a 10x/0.30 Plan - NEOFLUAR objective. The cell was placed
between crossed polarizers in order to detect birefringence
corresponding to the change in the order parameter. Starting at 1 bar we applied different pressures up to 1000 bar with steps of 200 bar steps. Pressurizing was performed with a rate of 100 bar/s. Pressure releases were performed from a sample in the nematic phase that had been equilibrated at a pressure of 1000 bar for about 1 h.
Consequently, we decreased the pressure starting from this equilibrated sample inducing the
nematic-isotropic phase transitions. After changing the
pressure, sequences of images were taken using a CarlZeiss Axiocam
CCD colour camera and Zeiss acquisition software. The sequences of
images were taken with different time resolution: in the beginning
every 5 s and in the end every minute over a total time of 3
hours. Optical birefringence measurements were performed on a home made setup consisting
of a argon ion laser line at $514.5 nm$, a beam expander, two polarizers and a Coherent
Fieldmaster-GS power and energy meter. The detector was operated by
a self written LabVIEW application. The pressure quenches were
applied with the same protocol as described above, starting always at 1 bar. The detector registered the intensity of the through-going beam with a time resolution of 0.4 s over a total time of 1 hour. To acquire more information about kinetics of the nematic-isotropic transition, measurements of the forward transmission were performed. When a phase transition takes place, the sample becomes turbid which is indicated by a decrease in the transmission. Turbidity was probed with a He-Ne laser and a Coherent Fieldmaster-GS power and energy meter. Pressure releases were performed as described above. The transmission was detected and registered as in the birefringence measurements with the help of the LabVIEW software.

\subsection{Small Angle Light Scattering (SALS) under shear}\label{ssec:ExpShear}

A homebuild optical couette shear cell combined with a SALS-setup was used~\cite{Holmqvist05}. The shear cell consisted of a rotating inner cylinder with a diameter of 43 mm and a static outer cylinder with a diameter of 47 mm resulting in a gap width of 2 mm. The inner and outer cylinders were both made of optical grade glass. A 15 mW HeNe Laser (Melles Griot) operating at a wave length of 632 nm was used. To ensure that the laser beam went through only one gap, it was directed through the center of the rotational axis of the inner cylinder. In the rotating cylinder, the beam was directed along the radial direction with a prism. Scattered intensities were projected with a lens directly on to the chip of a Peltier cooled 12-bit CCD camera, with 582x782 pixels (Princeton Instruments, microMAX). The scattering angles on the chip were calibrated by placing a known grid (PAT 13 Heptagon) in the scattering volume.

The fd-virus solution was always pre-sheared at $100 \; s^{-1}$ and quenched to zero shear rate at $t=0$, at which time the registration of the SALS patterns started with a rate of about two frames per second. Immediately after cessation of flow the rods will on average be oriented along the flow direction. Thus initially the system will be in a homogenous flow-induced nematic state with a well defined 'director' $\widehat{n}$. The effect of the pre shearing was checked for a number of samples by reducing the pre-shearing to $50 \; s^{-1}$. No difference on the result could be found and we therefore kept the protocol of a pre-shearing rate of $100 \; s^{-1}$.

\begin{figure}
\includegraphics[width=.405\textwidth]{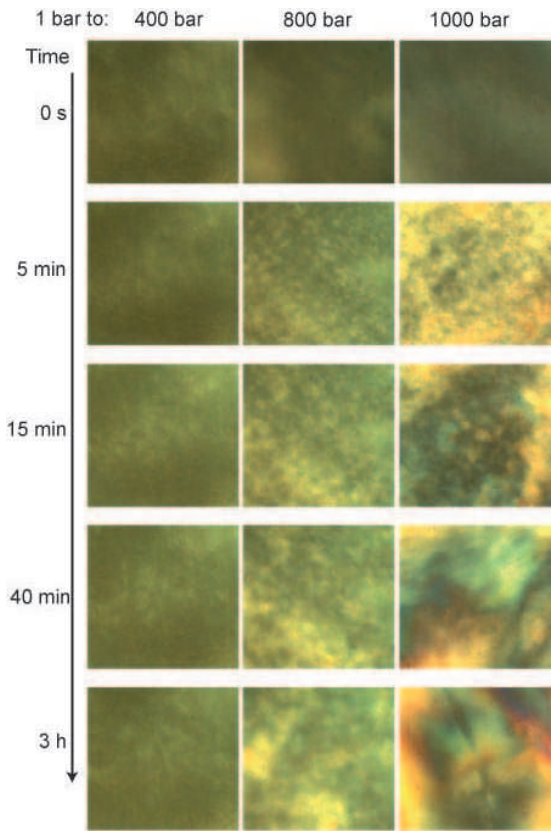}
\caption{Sequence of polarization microscopy image after an increase in pressure, starting at $1\;bar$. The final pressure is a measure of the concentration. Bright regions with a higher orientational order parameter appear at about 15 minutes after a quench to 400 bar  indicative of a nucleation-growth mechanism (middle left). For a final pressure of 1000 bar  phase separation sets in immediately (top right), indicative of spinodal decomposition,  while after 40 minutes a full nematic phase is formed (bottom right).} \label{fig:polaimgup}
\end{figure}

\begin{figure}
\includegraphics[width=.45\textwidth]{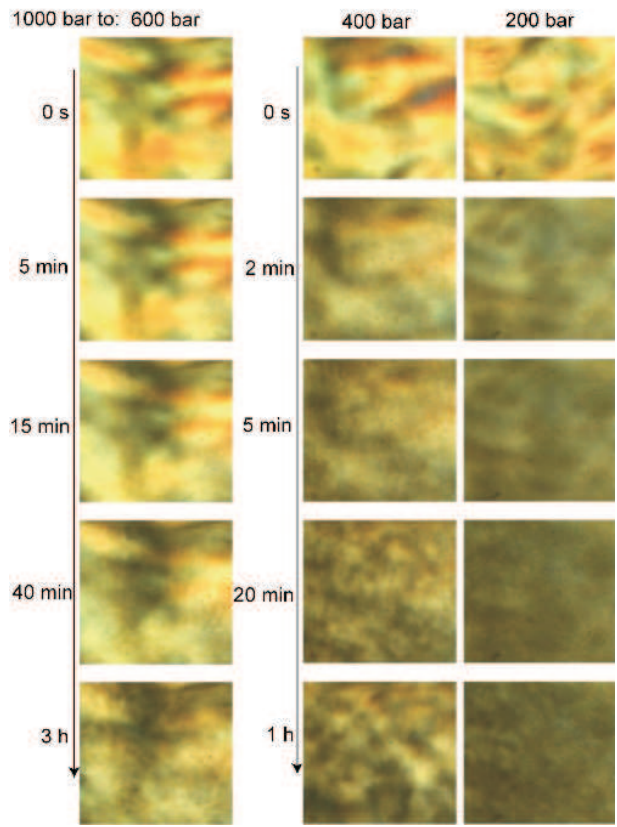}
\caption{As Fig. \ref{fig:polaimgup} but now starting at $1000\;bar$. Nucleation-and-growth events can be seen for a final pressure of 600 bar after 15 minutes (middle left). For a final pressure of 200 bar phase separation sets in immediately (top right), indicative of spinodal decomposition, but also the total intensity is much reduced, indicative of a lower ordering. After 1 hour a full isotropic phase is formed (bottom right).} \label{fig:polaimgdown}
\end{figure}

\section{Results}\label{sec:Results}

\subsection{Concentration quenches using pressure}\label{ssec:ResPres}

\begin{figure}
\includegraphics[width=.3\textwidth]{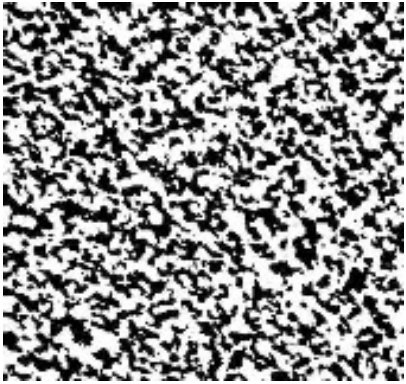}
\caption{Filtered and
binarized micrographs of \emph{fd }virus after the pressure jump from 0 to
1000 bar taken at 50 seconds after the quench. Growing
interconnected structures are a clear proof of spinodal decomposition
taking place.} \label{fig:spinodaldecomposition}
\end{figure}

Sequences of micrographs taken after pressure quenches for different time delays are gathered in Fig. \ref{fig:polaimgup} and \ref{fig:polaimgdown}.  The starting pressure for  Fig. \ref{fig:polaimgup} was 1 bar so that initially the system was in the isotropic phase. The varied depth of the quenches allowed for exploration of different regions on the phase diagram. For the 200 bar pressure quench the images stay dark throughout the whole experiment (data not shown). For a quench to400 bar the total intensity only starts to increase after about 5 minutes. At longer times brighter regions, indicative of a finite order parameter, are visible due to nucleation-and-growth. The possible presence of tactoids could not be observed, because a 10x objective was used. An induction time $\ti$ for phase separation to set in is characteristic for the meta-stable states. For a quench to 800 bar an almost instant change in the intensity is observed along with homogeneous structure formation. This observation suggests that the I-N spinodal is located in the immediate vicinity of this applied pressure. In the last sequence corresponding to the deepest quench of 1000 bar one can see the full transition from isotropic to nematic phase through spinodal decomposition. First, the increase in intensity started immediately after the quench, i.e. $\ti=0$. Second, the initial early stages of the transition exhibited morphology characteristic for spinodal decomposition - interconnected, labyrinth-like structures spanning through the whole sample. Figure \ref{fig:spinodaldecomposition} shows a thresholded picture taken at 50 seconds after this quench. Typical spinodal decomposition morphology is evident.  The last two micrographs of the 1000 bar quench show that the system develops into a fully nematic phase indicated by large homogenous regions of the same colour that only reorient with time.  Thus the I-N spinodal $\Csin$ is located in the very proximity of the N-I binodal $\Cbni$.

Fig. \ref{fig:polaimgdown} depicts the nematic-isotropic phase transition after the pressure quenches from 1000 bar to the indicated pressures. The difference in the morphology for the shallowest quench from 1000 to 800 bar are subtle and changes are only visible 3 h after quenching (data not shown). For the quench from 1000 to 600 bar the changes are more pronounced. After about 20 minutes grainy structures are formed,  characteristic for a phase separation via nucleation and growth. Such structures can still be seen in the last image of the sequence though the overall intensity decreased substantially. The image taken two minutes after the quench from 1000 to 400 bar shows that the overall intensity has decreased, indicative of a lower ordering, while already some biphasic dark and white structures seem to be visible. The first image after the deepest quench, from 1000 to 200 bar, shows that phase separation sets in immediately (top right), indicative of spinodal decomposition, while the total intensity is much reduced. After 1 hour a full isotropic phase is formed (bottom right) showing that a full phase transition took place. A more conclusive location of the spinodal and binodal points can be obtained with turbidity and birefringence measurements as presented in Sec. \ref{ssec:ResSpBi}.

Since with 1000 bar we approach the limit of the maximum applicable pressure, we cannot use this technique to study dispersions of rod-polymers mixtures. For such dispersions the width of the biphasic region (see Fig. \ref{fig:PhaseDex}) cannot anymore be bridged. For this reason we use shear flow to induce a flow stabilized nematic phase followed by cessation of shear flow.

\begin{figure}
\includegraphics[width=.45\textwidth]{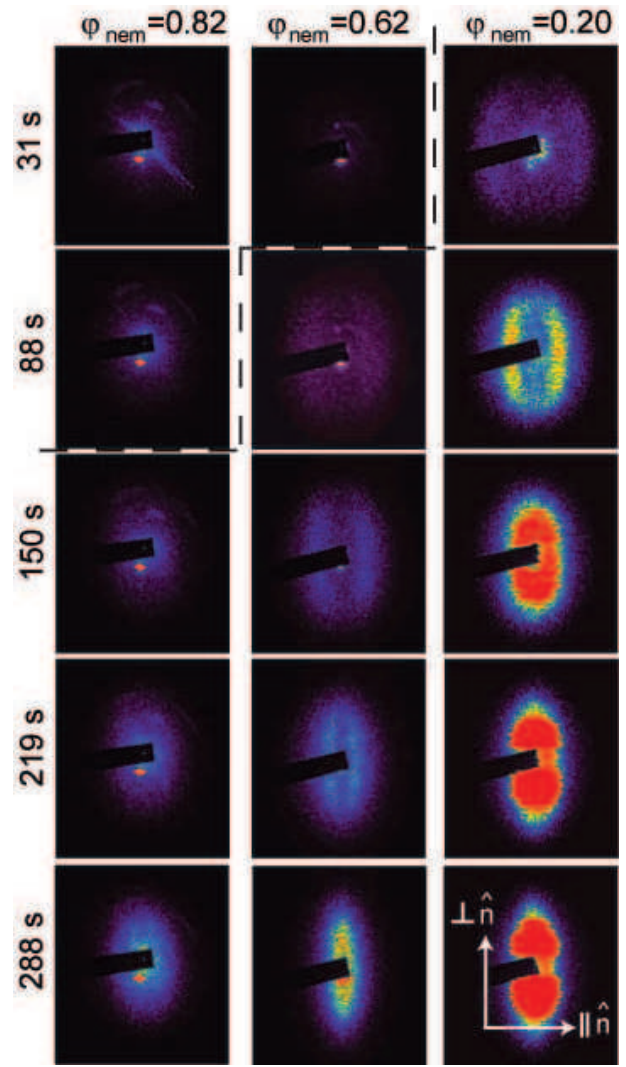}
\caption{Development of the scattering pattern for $\fn=$0.20, 0.62 and 0.82, after a shear rate quench from $100$ to $0\;s^{-1}$ for the high dextran concentration. For $\fn=0.20$ spinodal decomposition immediately sets in as can be concluded from the faint ring at 31 s (upper right), while for $\fn=0.82$ nucleation of structure can only be observed after about 150 s (middle left). The indicution time for structure formation is indicated by the dashed line.} \label{fig:salsimg}
\end{figure}

\subsection{Orientation quenches using cessation of shear flow}\label{ssec:ResShear}

\begin{figure}
\includegraphics[width=.4\textwidth]{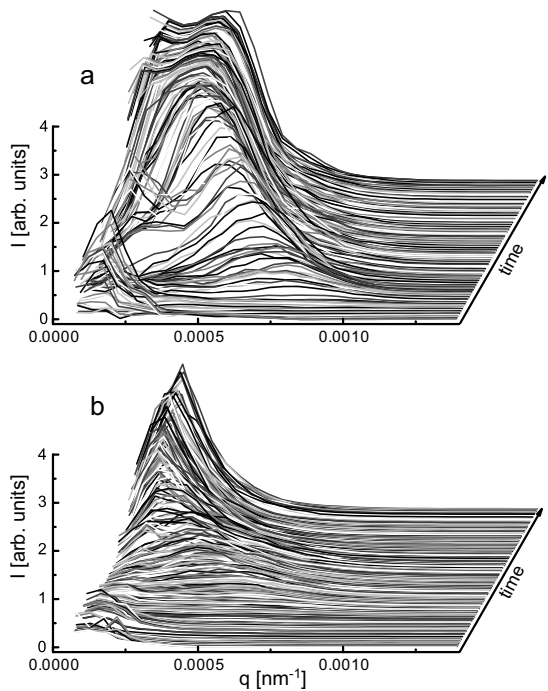}
\caption{Development of the scattering along the flow direction for $\fn=0.20$ (a) and  $\fn=0.62$ (b) as deduced from the scattering patterns in Fig. \ref{fig:salsimg}. The total time is three and a half minutes in (a) and six minutes in (b).} \label{fig:Ivsq}
\end{figure}

The development of the scattering pattern after a quench in the shear rate can be seen in Fig. \ref{fig:salsimg} for three different  $\fn$ : 0.20, 0.62 and 0.82. The time at which the first detectable  scattering structure appears increases with increasing  $\fn$, as indicated by the dashed line in Fig. \ref{fig:salsimg}. For the highest concentration this is the case only after about 150 seconds. For this concentration it is difficult to identify specific features. For the lower two concentrations the scatter pattern occurs as two slightly bend lines perpendicular to the director ($\widehat{x}$). In the early stage these lines scatter weakly but with time the intensity increases, the peak sharpens and the position moves to smaller wave vectors. The scattering patterns are isotropic in the initial stage but become increasingly more asymmetric with time. This indicates that the formed objects are increasingly more elongated and oriented along the director. To quantify the development of the formed biphasic structure we extract intensity profiles along  and perpendicular to the director. The time development of the scattering along the director is shown in Fig. \ref{fig:Ivsq}. Here the formation of the structure peak and both the movement to lower q-values and the intensity increase of this peak are clearly seen for the two lowest concentrations,  $\fn$, where it should be mentioned that for the lowest  $\fn$ the peak develops  immediately after cessation of flow, indicative of spinodal decomposition. At the highest concentration the structure peak can only be distinguished at longer times, characteristic for nucleation-and-growth. Fig. \ref{fig:Ivsq} and similar plots taken for other concentrations will be used in the following sections as a base to deduce  the induction time for structure formation.

\begin{figure}
\includegraphics[width=.4\textwidth]{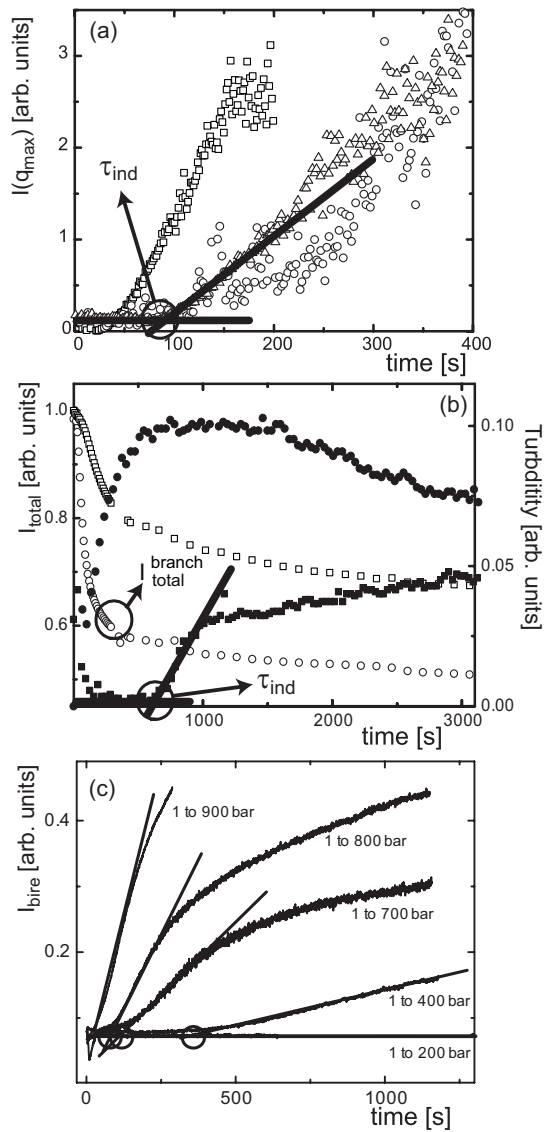}
\caption{(a) The intensity at the peak of the scattering pattern, $I(q_{\text{max}})$, after cessation of shear flow for $\fn=0.30$ ($\Box$), $\fn=0.62$ ($\circ$) and $\fn=0.82$ ($\vartriangle$) at the high dextran concentration to study the N-I transition. (b) Turbidity (solid symbols)and total intensity of the polarization micrographs (open symbols)  after a pressure drop (cubes: 400 bar, bullets : 200 bar) to study the N-I transition. The total intensity of the polarization micrographs after the initial decay, $I_{\text{total}}^{\text{branch}}$, is also indicated. (c) Birefringence intensity after an increase in pressure to study the I-N transition. The lines in the plots indicate the extrapolation to determine the induction time after a shear rate or pressure quench. } \label{fig:IvsTime}
\end{figure}

\subsection{Spinodal and binodal points}\label{ssec:ResSpBi}

Figures \ref{fig:polaimgup},\ref{fig:polaimgdown} and \ref{fig:salsimg} all show that after the quench phase separation sets in immediately or after some induction time $\ti$, depending on the depth of the quench. As explained in the introduction, in order to locate the spinodal and binodal points we have to find at what concentration $\ti$ for the formation of nematic (for the I-N transition) or isotropic structures (for the N-I transition) goes either to zero or to infinity.

\begin{figure}
\includegraphics[width=.45\textwidth]{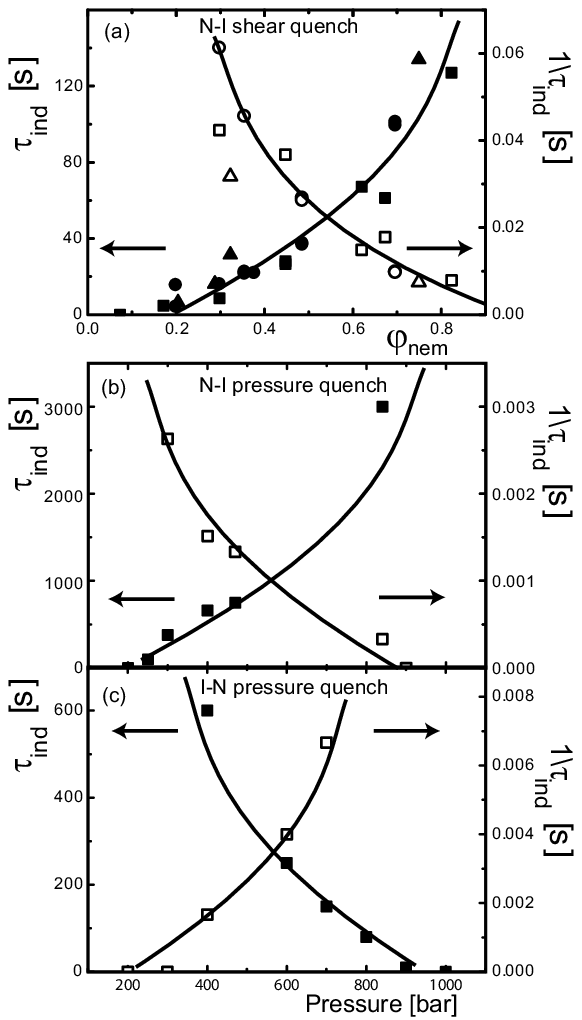}
\caption{Induction times (filled symbols) and inverse induction times (open symbols) starting with an initial nematic phase (a,b) to determine $\Cbni$ and $\Csni$ and with an initial isotropic phase (c) to determine $\Cbin$ and $\Csin$; (a) as a function of $\fn$  for three dextran concentrations (high ($\Box$), middle ($\circ$) and low ($\vartriangle$)) after pre-shearing; (b,c) as a function of the final pressure after an initial pressure of 1000 bar and 1 bar, respectively. The lines are a guide to the eye.} \label{fig:IndVsConc}
\end{figure}

To determine $\ti$ for quenches in the orientation we plot the intensity of the peak in the scattering pattern, see Fig. \ref{fig:Ivsq}, as a function of time. Fig. \ref{fig:IvsTime}a shows this time development for three different $\fn$. The induction $\ti$ is now obtained by extrapolating the linear intensity increase to zero intensity. This procedure was repeated for different attraction strength, i.e. different dextran concentrations. The result is shown in Fig. \ref{fig:IndVsConc}a, where  respectively $\ti$ and $1/\ti$ are plotted vs.  $\fn$ for the three attractions. The N-I spinodal $\Csni$ and binodal $\Cbni$ points are determined by extrapolating $\ti$ and  $1/\ti$ to zero, respectively. Interestingly, the curves of  $\ti$ and  $1/\ti$ for the different attractions  overlap if we scale the concentration relative to the phase boundaries. As a consequence the location of $\Csni$ is independent of the attraction and is found to be $\fn\;=\;0.25$. More striking even is the observation that also the absolute induction times are only effected by the relative distance from the phase boundary $\fn$ and not the absolute \emph{fd }concentration or the attraction, i.e. the dextran concentration. Of course the phase boundaries can also be determined by measuring the concentration of the two separated phases or visual observation, considering that no scattering pattern should occur for the homogeneous phases. The deviation between the different methods is less than 5\%. This means that the high concentration phase boundaries can be determined in three different ways and the low concentration phase boundary in two ways.

The spinodal decomposition is characterized not only by the fact that it immediately sets in, but also by the morphology of the formed structure, which should be bicontinuous. It has been shown that the transition in the morphology is quite smooth~\cite{Lettinga05c,vanbruggen99a}. This is also exemplified by Fig. \ref{fig:salsimg} and Fig. \ref{fig:Ivsq}b for $\fn=0.62$. A clear ring structure is observed, exemplary for spinodal structures, but also for this concentration an induction time is observed. Thus the phase separation has already spinodal characteristics, but clearly the system is still in the meta-stable region. Therefore the only reliable way to determine the spinodal point is by using the induction time.

The kinetics of the phase separation for samples without polymer was studied using the pressure quench. The reason is that with this technique we access both the N-I \emph{and} the I-N transitions. As described in section \ref{ssec:ExpPres} the location of these points could be estimated from microscopy experiments. The interpretation of the micrographs is, however, not straightforward. Quenches starting from an initial nematic phase at 1000 bar show that the overall intensity decreases as well as that structures are formed. To separate the two effects we additionally performed turbidity measurements,  since turbidity is a measure of the biphasic structure that is created in the sample during phase separation. In Fig. \ref{fig:IvsTime}b the responses of the turbidity and the total integrated intensity $I_{total}$ of the polarization micrographs are plotted for quenches to 400 and 200 bar. For the quench to 400 bar clearly an induction time is observed in the turbidity, while the birefringence, which dominates the intensity of the polarization microscopy, shows a fast and a slow decay. Since the flow induced nematic phase has a higher order parameter than the metastable nematic branch (see Fig. \ref{fig:PhaseScheme}) the nematic phase will first relax to this branch, which explains the initial fast decay of the birefringence to  $I_{\text{total}}^{\text{branch}}$. Turbidity is not sensitive for this process, because structure formation is not involved in this process. Hence we can conclude that phase separation sets in only after an induction time of about 650 seconds. For the quench to 200 bar changes in birefringence and turbidity  set in immediately so $\ti=0$. Since we know from microscopy that the final stage for 200 bar is fully isotropic, the down turn in the turbidity after half an hour can be interpreted as the disappearance of biphasic structure. Before reaching the fully isotropic phase (low turbidity) the system undergoes a phase transition (high turbidity). When pressurizing an initially isotropic sample then the increase of birefringence can only be caused by the formation of the nematic phase. Thus the induction times for the formation of the nematic phase can be obtained from the birefringence responses as plotted in Fig. \ref{fig:IvsTime} c.

\begin{figure}
\includegraphics[width=.4\textwidth]{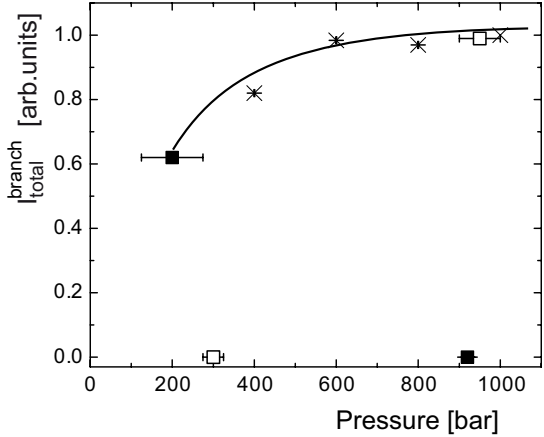}
\caption{The experimental equivalent of the bifurcation diagram. The pressures corresponding to $\Csni$ and $\Csin$ are determined from Fig. \ref{fig:IvsTime}b and c as the pressures where $\ti\rightarrow 0$ (solid symbols). The pressures corresponding to $\Cbni$ and $\Cbin$ are given by the pressures where $1/\ti\rightarrow 0$ (open symbols). The total intensity of the polarization micrographs after the initial decay,  $I_{\text{total}}^{\text{branch}}$, (see Fig \ref{fig:IvsTime}b) is taken as a measure of the orientational order parameter (stars). The line is a guide to the eye.} \label{fig:PhaseOut}
\end{figure}

The induction times and the reverse induction times are plotted in Fig. \ref{fig:IndVsConc} b and c for the turbidity and birefringence, respectively. The pressures corresponding to $\Csni$ and $\Csin$ are determined from these two figures as the pressures where $\ti\rightarrow 0$, while the pressures corresponding to $\Cbni$ and $\Cbin$ are given by the pressures where $1/\ti\rightarrow 0$. We can now construct the experimental equivalent, Fig. \ref{fig:PhaseOut}, of the theoretical bifurcation diagram, Fig. \ref{fig:PhaseScheme}, using pressure as a measure of concentration and  the total intensity of the micrographs after the initial decay as a measure of the orientational order parameter (Fig. \ref{fig:IvsTime}b). Note that the error bar in the pressure is determined from the uncertainty in the extrapolation to $\ti\rightarrow 0$ and $1/\ti\rightarrow 0$. Another source of error could also be the exact concentration of the rods, since the data were taken on different batches. This error would show up as a shift of the entire curve that is obtained from the used batch. This could explain why $\Csni$ is somewhat smaller than $\Cbin$, which is in principle not possible.

\begin{figure}
\includegraphics[width=.4\textwidth]{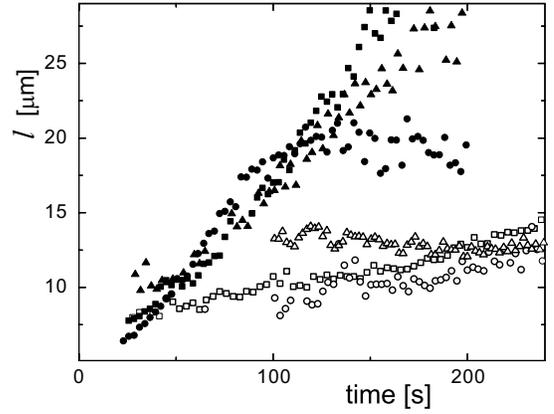}
\caption{Time development of the structure size parallel (solid symbols) and perpendicular (open symbols) to the director, for three dextran concentrations at a scaled volume fraction of $\fn=0.30$. Symbols as in Fig. \ref{fig:IndVsConc}a.} \label{fig:SizeVsTime}
\end{figure}

\subsection{Growth rates}\label{ssec:Grow}

\begin{figure}
\includegraphics[width=.4\textwidth]{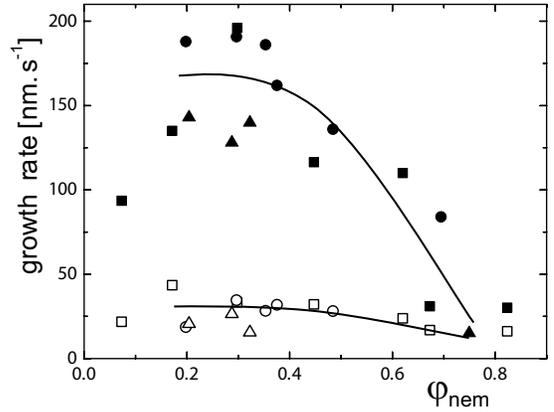}
\caption{The growth rate vs. $\fn$ parallel (open symbols) and perpendicular (solid symbols) to the director for three  dextran concentrations. Symbols as in Fig. \ref{fig:IndVsConc}a.} \label{fig:RateVsPhi}
\end{figure}

The average size of the structures that are formed during the phase separation can be deduced from the location of the peak of the scattered intensity. The resulting size parallel and  perpendicular to the director are plotted in Fig. \ref{fig:SizeVsTime} as a function of time. Surprisingly the size of the formed structures as well as the rate with which they grow do not depend on the attraction. The structure formed in the early stage is isotropic, while the structure growth is anisotropic: the structures are growing faster along the director so that the structure becomes anisotropic with time but the anisotropy in the structure does not seem to increase beyond an aspect ratio of 2. This can also be appreciated from Fig. \ref{fig:RateVsPhi}, where the extracted growth rates are plotted against $\fn$ for all attractions in. The only factor that affects the growth rate is the distance from the phase boundaries. As expected the growth rates go to zero at $\Cbni$ while the growth rate seems to go to a plateau value around $\fn=0.25$, which is, as we have seen in the last paragraph, the spinodal point $\Csni$.

\begin{figure}
\includegraphics[width=.4\textwidth]{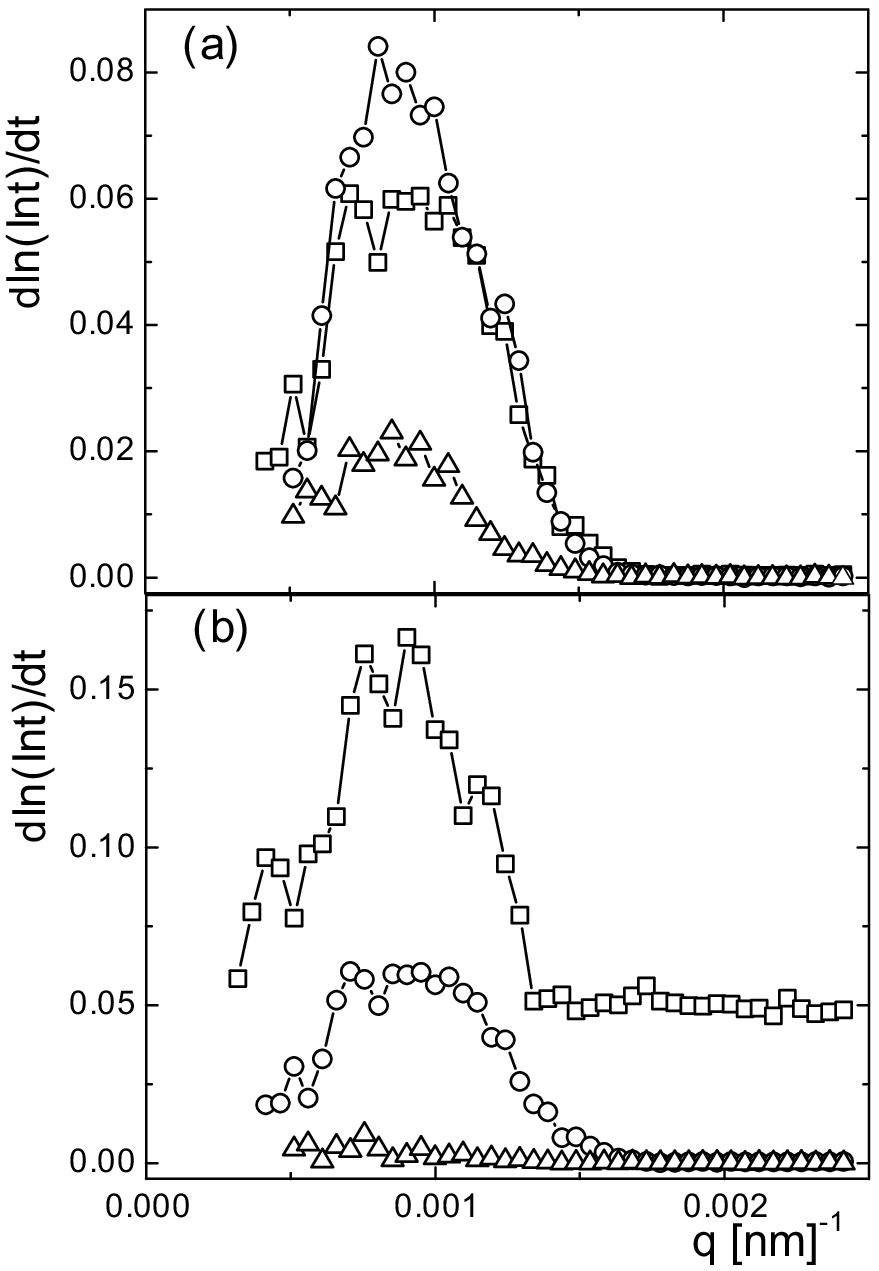}
\caption{  ln(Int)/dt vs. $q$ for a) three different $\fn$: $\fn=0.20$ ($\Box$), $\fn=0.30$ ($\circ$) and $\fn=0.70$ ($\vartriangle$) for the middle dextran concentration and b) three dextran concentrations at  $\fn=0.2$. Symbols as in Fig. \ref{fig:IndVsConc}a.} \label{fig:RateVsQ}
\end{figure}

To access the very early changes, below 25 s, we look at the logarithm of the intensity $\ln(I)$ with time, at different the wave vectors $q$. In this way we probe the length scale of the density fluctuations which has initially the highest probability to grow. This investigation is only done along the director since for the neutral direction the signal to noise ratio is too low. In Fig. \ref{fig:RateVsQ}a $d\ln(I)/dt$ is plotted as a function of $q$ for three different $\fn$ for sample B. $d\ln(I)/dt$ peaks at the same $q$ vector for all $\fn$, i.e. independent if it is in the unstable region or far in the meta-stable region. The length scale over which the system grows fastest turns out to be of the same size as the objects which have just formed at the induction time, i.e. the critical nuclei. For the nucleation-growth regime this means that in the first time window new nuclei are formed continuously with a well defined size, which start growing at the induction time.
In Fig. \ref{fig:RateVsQ}b $d\ln(I)/dt$ vs. $q$ is plotted for the three different attractions at a volume fraction of $\fn=0.2$, i.e. for a spinodal decomposing sample. Again the maximum $q$ of the growth is the same for all attractions, as was found for the induction time. On the other hand the amplitudes are different. This amplitude different is due to the change in contrast different between the isotropic and nematic. With increasing dextran concentration the width of the biphasic region increases and thus the density and contrast between the two phases. If we now combine this observation with the observation that both the growth rate of the objects and the induction time do not change with added attraction it is evident that what by eye seems to be a faster phase separation with more dextran added only is a optical effect due to the increasing in contrast.

\section{Discussion}

The main goal of our investigations was to study the extent of the supercooled or superheated regime for dispersions of attractive rods and how it depends on the strength of the attraction. This dependence is given in Fig. \ref{fig:PhaseDex}. The location of the N-I spinodal point almost does not change in the range of attractions studied here. This holds both for the absolute concentration of \emph{fd }virus (Fig. \ref{fig:PhaseDex}) as well as the fraction of the coexisting nematic phase $\fn$ (Fig. \ref{fig:IndVsConc}a), at least within the experimental error. These observations are confirmed by theory, where it is found that both spinodal lines as a function of the absolute rod concentration are insensitive of the attraction (Fig. \ref{fig:theo1}a). The N-I spinodal line has a weak dependence on the attraction when plotted as a function of $\fn$ (Fig. \ref{fig:theo1}b).

Using $\fn$ as the scaled concentration we observed that not only the location of the spinodal is insensitive for the attraction, but also the size of the critical nuclei that are formed in the meta stable region, Fig. \ref{fig:SizeVsTime}, the induction after which they start to grow, Fig. \ref{fig:IndVsConc}a, and the rate with which they grow, Fig. \ref{fig:SizeVsTime}. This is surprising since in classical nucleation theory these parameters depend on the difference in the chemical potential between the homogeneous and demixed state and the interfacial tension between the two formed phases\cite{Debenedetti96}. These thermodynamic parameters are expected to be quite different, considering that the width of the biphasic region for the highest used dextran concentration has increased with an order of magnitude, see Fig. \ref{fig:PhaseDex}. Given the fact, however, that the relative location between $\Cbni$ and $\Csni$ set the length and time scales, there must be another parameter that determines the nucleation barrier. To gain theoretical understanding of this problem would require a full dynamical density functional approach, including spatial inhomogeneities ~\cite{Dhont05},  to access the evolution of the microstructure over time. Using Monte Carlo simulations Schilling et al.\cite{Schilling07} did not find an increased orientation correlation for polymer volume fraction comparable to those used in the experiments. If the initial stage of phase separation is dominated by collective rotational diffusion, then both experiments and the simulation~\cite{Schilling07} hint that depletion interactions do not lead to a stronger preference of rods to align. Simulations by Cuetos et al.\cite{Cuetos07} confirm that the aspect ratio of critical nucleus is less than two, independent of supersaturation. The size of the critical nuclei is however in the order of one rod length, whereas our measurements indicate that the critical nucleus is about 7 rods long independent of the depth of the quench or the attraction strength. This is most clearly shown in Fig. \ref{fig:RateVsQ}, because this figure plots at what density wavelength the intensity growths fastest. Combination of simulations and theory does show that the anisotropy in the surface tension due to the planar anchoring of rods at the interface plays an important role in the formation of a critical nucleus~\cite{Cuetos08}. If this effect dominates the kinetics, it is understandable that  attraction between the rods is less important than for example the aspect ratio of the rod. Comparing experiments and simulations one should realize that not only the aspect ratio of the viruses is an order of magnitude bigger than those used in simulations, but also that we consider in Fig. \ref{fig:RateVsQ} the N-I and not the I-N transition.

The only parameter that is influenced by the attraction is the rate at which the gradient in the density grows, as plotted in Fig. \ref{fig:RateVsQ}b, which is due to the increasing width of the biphasic region. All samples in this plot are quenched into the unstable region and therefore undergo spinodal decomposition. The curve for the highest attraction, i.e. the  highest polymer concentration, shows a clear peak. This is typical for spinodal decomposition that is dominated by translational diffusion~\cite{Matsuyama00,Winters00}.  At the middle concentration there is still a clear peak, but it is not obvious wether the demixing rate goes to zero for $q\rightarrow 0$ or not. For the lowest concentration of dextran the data are too noisy to draw any conclusions. Our results are somewhat different from earlier experiments performed at $13\;mg/ml$ of dextran, which hinted that the growth rate does not go to zero for $q\rightarrow 0$ ~\cite{Lettinga05c}. Also the binodal lines found in this paper have sharper features as compared to those published in Ref.~\cite{Dogic04a}. Differences in the poly-dispersity of the dextran that is used could explain these discrepancies. In the latter paper also a significantly large part of the phase diagram was covered. The reason for the somewhat limited range of attraction studied here is that higher concentrations of dextran would lead to a too high turbidity of the sample and multiple scattering, corrupting the reliability of the measurements. This effect could explain the fact that for the high polymer concentrations in Fig. \ref{fig:RateVsQ}b  a finite growth rate is found at high $q$ values. Concerning the time dependence of the structure growth we found within experimental error a power law of around one, see the linear dependence in Fig. \ref{fig:SizeVsTime}a, whereas theory predicts a lower power dependence\cite{Dhont05}. Possibly we are restricted to the very initial stage of the phase separation process. The sizes of the coalescing structure formed at later times are too big, so that the scattered light hits the beam stop. This problem does not occur when using microscopy as in Ref. ~\cite{Oakes07,Chowdhury90}.

With the pressure quench we accessed both the N-I and the I-N transition. As for the N-I transition we observed that also $\Csin$ is located in the proximity of $\Cbni$. With this we confirm for the first time the theoretical prediction done in Ref. ~\cite{Kayser78}. What these experiments also show is that for the deepest quenches the initially homogeneous single phase (I or N) undergoes a local phase separation before it completely turns in to the new single phase. This mechanism confirms similar observations in computer simulations\cite{Cuetos08} on initially isotropic hard spherocylinder-polymer mixtures. The fact that we can reach the full nematic state from the isotropic state with the pressure quench is striking, since it is known that the concentration difference between the two binodal points is 10 \%, whereas with a pressure quench to 1000 bar the water is compressed only 5 \%. Here it is important to note that we are comparing the isotropic phase at 1 bar with the nematic phase at 1000 bar. Assuming that the width of the biphasic region does not change with pressure, it would mean that at high pressure the I-N transition sets in at lower concentrations. It is known that the location of the I-N transition is temperature dependent. This dependency could be linked to the temperature dependence of the flexibility of the {\it fd} virus~\cite{Tang96}. Similarly, the features of {\it fd} virus could changes at high pressures. This change cannot be an irreversible process like denaturation, since the phase transitions are completely reversible. Further experiments are needed to show if is a specific feature like the flexibility of rod-like viruses that changes with pressure. Alternatively, more general features like the Debye double layer of charged colloids could be pressure dependent.

\section{Conclusions}

We have studied the behavior of supersaturated dispersions of rod-like viruses. Superheated nematic dispersions were prepared by first applying a strong shear flow. Cessation of the shear flow at $t=0$ renders the nematic phase meta-stable or unstable depending on the concentration, see Fig. \ref{fig:PhaseScheme}. We probed the structure formation using Small Angle Light Scattering. With the analysis of the scattering patterns we could access the induction time for structure formation $\ti$, the size of the critical nucleus and the growth rate. These parameters were measured over a broad range of attractions as induced by the addition of dextran as a depletion agent. We found that the N-I spinodal point, i.e. the concentration where $\ti \rightarrow 0$, is independent of the attraction, which was confirmed by theoretical calculations. Interestingly, also the absolute induction time, the critical nucleus and the growth rate are insensitive of the attraction, when the concentration is scaled to the distance to the phase boundaries as given by $\fn$. This observation hints that concepts of classical nucleation theory are insufficient to understand nucleation processes in anisotropic fluids. We also applied pressure quenches on dispersions of rods without added polymer, thus supercooling or superheating the system. The pressure quenches were deep enough to induce a complete phase transition from the isotropic to the nematic phase and vice versa, which takes place initially via phase separation. As a consequence both the N-I and I-N spinodal could be accessed. By a combination of polarization microscopy, birefringence and turbidity measurements we were able to construct a first experimental analogue of the bifurcation diagram of Kayser and Ravech\'{e}~\cite{Kayser78}.

\section{acknowledgments}

This work was performed within the framework of the Transregio SFB TR6, 'Physics of
colloidal dispersions in external fields'. The authors thank  R. Tuinier, A. Patkowski and J. Dhont for stimulating discussions. D. Kleshchanok is thanked for critical reading of the manuscript. R. Vavrin is thanked for technical assistance with the turbidity measurements.

\end{document}